# Security Evaluation for Mail Distribution Systems

A. S. Rizopoulos[1], D. N. Kallergis[1], G. N. Prezerakos[1]

[1] Dpt. of Electronic Computer Systems

Technological Educational Institute (TEI) of Piraeus

P. Ralli & Thivon, 122 44, Aigaleo, Greece

E-mail : a.rizopoulos@s2e.teipir.gr, {d.kallergis, prezerak}@teipir.gr

**ABSTRACT**

The security evaluation for Mail Distribution Systems focuses on certification and reliability of sensitive data between mail servers. The need to certify the information conveyed is a result of known weaknesses in the simple mail transfer protocol (SMTP). The most important consequence of these weaknesses is the possibility to mislead the recipient, which is achieved via spam (especially email spoofing). Email spoofing refers to alterations in the headers and/or the content of the message. Therefore, the authenticity of the message is compromised. Unfortunately, the broken link between certification and reliability of the information is unsolicited email (spam).

Unlike the current practice of estimating the cost of spam, which prompts organizations to purchase and maintain appropriate anti-spam software, our approach offers an alternative perspective of the economic and moral consequences of unsolicited mail. The financial data provided in this paper show that spam is a major contributor to the financial and production cost of an organization, necessitating further attention. Additionally, this paper highlights the importance and severity of the weaknesses of the SMTP protocol, which can be exploited even with the use of simple applications incorporated within most commonly used Operating Systems (e.g. Telnet).

As a consequence of these drawbacks Mail Distribution Systems need to be appropriate configured so as to provide the necessary security services to the users.

*Keywords: Mail Server Security, Mail Distribution System Evaluation, Cost of Spam, Email spoofing.*

## 1. Introduction

Information Society is stated as the one in which the creation, distribution, diffusion, use, integration and manipulation of information is a significant financial, political, and cultural activity. In this society, messages sent or received electronically by its citizens are said to be the cornerstone of its structure.

Digital citizens are bound to increasingly trust the intercessors of their communication process, which in our case are machines that distribute electronic messages. As any machinery constructed until now, these systems have some weaknesses, which in some cases may cause significant damage by means of financial, as well as ethical, impact.

In this paper, we address this problem, focusing on economy, as well as ethics. We propose a solution that focuses on increasing the amount of trust felt by users toward email messages. We show that certain weaknesses of the SMTP protocol can be exploited by a simple application such as Telnet which is incorporated within most operating systems. As a solution, we propose a user authorization procedure so that an unauthorized user cannot send emails. In this way, we manage to appropriately configure our Mail Distribution System in order to provide the necessary security services to the users.

## 2. Taxonomy of threat models

Several frameworks for threat modeling have been proposed, e.g. the OCTAVE model from CERT [1] and the STRIDE/DREAD methodology from Microsoft [2][3].Threat modeling is used for the identification and prioritization of security vulnerabilities of a system. A classification process based on threat modeling is advantageous because it offers increased security. Microsoft's STRIDE model is used to categorize different threat types. STRIDE is an acronym of the following concepts.
- **Spoofing.** Attempt to gain access to a system using a forged identity. A compromised system would have access control vulnerability.
- **Tampering.** Manipulation of data during communication through the network. The integrity of the data is threatened.
- **Repudiation.** Denial of participation in a transaction. The availability of a resource is threatened.
- **Information disclosure.** Unwanted exposure and loss of confidentiality of private data.
- **Denial of service.** Attack on system availability through the depletion of system resources.
- **Elevation of privilege.** A user with limited privileges assumes the identity of a privileged user to gain access to an application. The confidentiality, integrity, and availability of a resource are threatened.

DREAD is a risk-calculating mechanism. Each letter of the acronym stands for a threat attribute. Each of the attributes is ranked in a scale of 1 to 10 with 1 being the lowest and 10 being the highest rating. These attributes are:
- **Damage potential.** The damage that will be done if the vulnerability is exploited by the attacker.
- **Reproducibility.** The ease of repeatedly exploiting a vulnerability.
- **Exploitability.** The skill level required to exploit a vulnerability.
- **Affected users.** The parties by the exploitation of a vulnerability.
- **Discoverability.** The ease of exploration and discovery of a vulnerability.

In this paper, we focus on spoofing and information disclosure. Spoofing is the process by which spammers send a message, which is a false statement, or simply intended to mislead the recipients. It conforms to the term of ethical hacking, since it entails gaining unauthorized access to computer systems [4] - provided that no data is harmed, stolen or altered in this process. Phishing is the main spam technique for obtaining privileged information. It contains some of the attributes of spoofing, but its as main purpose is to compromise the privacy of the users. Therefore, ethical hacking can be used as a risk assessment technique, as proposed in [5].

## 3. Social and financial impacts

We believe that the user's trust toward the system is endangered when threats such as spam, phishing, email spoofing, etc. remain uncontrolled. We also believe that a human's trust towards the machinery is crucial, because it reflects his capabilities. In addition to trust, however, these threats can have other serious consequences (exacerbated by the increased adoption of digital communication), as illustrated by the following scenario: a forged email containing information on tax return issues is sent to thousands of citizens inviting them to provide some necessary documents. The consequences of such an event are fairly obvious. Apart from the impact on the governmental administration mechanism, there will also be financial costs to all involved parties.

Similarly, from an engineering perspective, the cost of spam is related to the consumption of network or system resources, as well as to user productivity. Users must spend time to evaluate messages, delete spam or look for "false positives", such as junk email messages that are not actually spam. Moreover, some costs arise from the types of spam and the way it affects the end-user. Every organization that provides a Mail Distribution System tries to limit or prevent the traffic of junk messages on its network. Thus, there is a constant need for additional equipment (hardware and/or software) in order to achieve better results. The table below outlines the main costs of unsolicited mail.

| Overhead | Transaction cost | Risks | Mechanical damage |
|---|---|---|---|
| Network bandwidth, storage need and development or acquisition of spam detection and prevention software. | The incremental cost of communications for each recipient individually along with the creation of a spam message that is multiplied by the number of recipients. | Possible legal move and/or public reactions, including claims and damages. | Implications for the community and/or communication channels which are "attacked" by spam |

**Table 1: General costs of spam**

Junk mail is the primary form of unsolicited communication because it presents virtually no cost to the sender. As mentioned before, in a productive environment, spam affects the economy of an organization and the productivity of its employees.

### 3.1. Case study

We used a free web-based application [6] to calculate the cost of spam on a public sector working environment. The input and output data of this case study are outlined in Table 2.

| | | | Total Corporate Cost of Spam | Cost of Spam for Each Employee |
|---|---|---|---|---|
| **Input data** | Number of employees with email accounts | | 680 | |
| | Number of workdays per year per employee | | 230 | |
| | Average hourly wage per employee (€) | | 15,00 | |
| | Average number of spam emails per day per employee | | 25 | |
| | Number of seconds wasted with each spam message | | 3 | |
| **Output data** | **Financial cost (€)** | Per year: | 48.875,00 | 71,88 |
| | | Per day: | 212,50 | 0,31 |
| | **Cost in terms of productivity *(days)*** | Per year: | 215,45 | 7,6 |

**Table 2: Case study: Results of spam cost calculator**

This is a typical paradigm of a public sector organization. Average wages, as well as workdays per year, have been used. It is concluded that approximately 50.000€ are wasted because of unsolicited email distribution, within a single year. Such an amount should be considered as critical and it is certainly one of the first targets when a "cut-off expenses" policy is about to be applied by the administration.

Other economists' estimations [7] refer that the cost for an entire country's economy is translated into more than 9 billion Euros per year.

## 4. Software apparatus

### 4.1. Breaking in

Spoofing exploits the two main reasons why SMTP is described as weak [8], [9]. These reasons are related to the user authentication process, which is taken advantage of by spammers. SMTP does not require user authentication; therefore, users have the opportunity to hide their identity. Moreover, every part of the message (headers, body, etc.) may be forged, resulting in the potential misrepresentation of the message by the recipient.

Especially today, most users' behavior is based on the hypothesis that if the source of the message is known (and therefore trusted), the content is reliable. Spammers can cause significant problems by exploiting this assumption. This method is featured in phishing and is achieved by the unsolicited mail.

More specifically, the sender needs to know at least two active email addresses and the domain name of the mail server in which the users' email accounts are stored. In this paper, the case of a mail server that is not working as an open relay is considered. The process of sending a falsified email is detailed below.

```
C¹   telnet smtp.mail.gr 25
S²   220 smtp.mail.gr M.T.A.
C    EHLO www.test.com
S    250-smtp.mail.gr
     250-PIPELINING
     250-SIZE 8192000
     250-ETRN
     250-STARTTLS
     250-AUTH LOGIN PLAIN
     250-AUTH=LOGIN PLAIN
     250-ENHANCEDSTATUSCODES
     250-8BITMIME
     250 DSN
C    MAIL FROM:<secr@mail.gr>       //sender address
S    250 2.1.5 Ok
C    RCPT TO:<professor@mail.gr>    //recipient address
S    250 2.1.5 Ok
C    DATA                           //start of message
     Date: Wed, 22 Jul 2009 13:56:45 +0300
     From: "Secretary" <secr@mail.gr>
     To: "Professors" <professor@mail.gr>
     Subject: Board of Examiners
     Reply-To: secr@mail.gr
     User-Agent: Webmail/0.2.0
     Content-Transfer-Encoding: 8bit
     Content-Type: text/plain; charset="UTF-8"
     You are invited to the Board of Examiners meeting
     scheduled for Thursday 16 September 2010 at 12.30
     p.m. at the department's council room.

C    .                              //end of message
S    250 2.0.0 Ok: queued as 492381B9295
C    QUIT
S    221 2.0.0 Bye
```
**Figure 1: SMTP session between a mail server and a mail client**

Initially, an SMTP session between the mail client and the mail server is established. When the "`HELO/EHLO`" command is executed, the mail server presents some public information about the functions it supports. The sender's email address is indicated by executing the SMTP command "`MAIL FROM:`". The recipient's address is indicated by running the "`RCPT TO:`" command. If no error occurs, the mail server confirms the validity of these email addresses. The initialization of the message data occurs with the execution of the "`DATA`" statement. The message data contains the headers and body of the email that will be sent. The end of the data segment is signified by the dot (`.`) and the `CR/LF` (carriage return/line feed) characters. Subsequently, the mail server responds with the message identification code. The SMTP session is terminated by the "QUIT" command. So, as a result of the sample code provided above, an email is sent to professor@mail.gr and the sender appears to be secr@mail.gr.

---
[1] "C" stands for mail client.
[2] "S" stands for mail server.

4.2 Our implementation

An efficient and robust solution to email spoofing is the use of User Authentication in combination with Real Time Blackhole Lists and Public Key Infrastructure. This implementation offers increased flexibility, as several tools and utilities exist that can render the Mail Distribution System even more secure.

The proposed method revolves around the authentication of users over an encrypted SMTP session. If the authentication process is successful, the message is composed and sent successfully. If the authentication process fails, an error message may be shown and the process is terminated.

We hosted our mail server on a machine with an Intel Dual Core 2.66Mhz processor, 3GB of RAM, 2 SATA disks with 250GB capacity each and a 100Mbps Internet connection. The server runs on Gentoo Minimal 2008 and the Mail Transfer Agent (MTA) software used is Postfix [10], following the directions and conclusions described in [11]. Spam Assassin [12] was used as the main anti-spam software, Courier-IMAP [13] as the IMAP/POP3 server, and Cyrus-SASL [14] as the authentication back-end.

Therefore, the SMTP session between the mail client and the mail server is as follows.

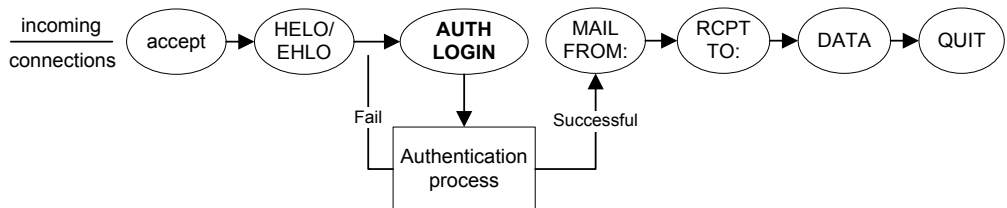

Figure 2: Sending process using authentication in SMTP session

After the SMTP connection is established, the client sends the "HELO/EHLO" command followed by its hostname or IP address. The server responds with its hostname and other information such as the supported message size, the supported protocols, etc. Subsequently, the mail client must send the "AUTH LOGIN" command to initiate the authentication procedure. Next, the username of the email account is sent in Base64 encoding. Afterwards, the user password must be sent, also in Base64 encoding. If the credentials sent match those stored in the user account, the process continues as described in Figure 1. In all others cases (if, for example, the client sends the "MAIL FROM" command before the "AUTH LOGIN" command), the mail server warns the client with a message such as "553 5.7.1<email address>: Sender address rejected: not logged in", stops the procedure, and waits for the proper command for a specified amount of time.

Using the method described above, we have prevented the unwanted behavior of unauthenticated users and the traffic of spoofed messages via our mail server.

For educational purposes, we ran some email spoofing tests on several SMTP servers. These tests took place in October 2009 and the results are presented in Table 3. The positive annotation within the "Secured" column explains that the vulnerability discovered, was fixed sometime after out trials. We refrain from providing the fully qualified hostname of these machines due to ethical concerns [15].

|  | Mail servers | Vulnerable | Secured |
|---|---|---|---|
| **Academic Institutes** | mail.xx.yy.zz | Yes | No |
|  | ulysses.xx.yy.zz | Yes | Yes |
|  | mail.xx.yy.zz | Yes | Yes |
|  | smtp.xx.yy.zz | No | Yes |
|  | mail.xx.yy | No | Yes |
|  | mail.xx.yy.zz | No | Yes |
| **Internet Service Providers** | dcmail01.xx.yy | Yes | No |
|  | mailgate.xx.yy | No | Yes |
| **Governmental Organizations** | mail.xx.yy | No | Yes |
|  | mail.xx.yy.zz | No | Yes |

Table 3: Vulnerable SMTP servers

## 5. Conclusion and future work

There is a constant need to ensure the authenticity and the reliability of any email, necessitating a firm grasp of related security risks. Thus, any Mail Distribution System needs to be configured so that it provides users with all the necessary services, with an emphasis on the reliability and the security of any communication process. Especially today, with email being the most commonly used communication platform, users must consider it as trustworthy and secure as possible.

In addition to the above, there are alternatives for reducing and/or eliminating email spoofing. A potential solution is an SMTP Proxy implementation. More commonly, these proxies are used in the integration of anti-spam techniques into MTAs. There are many advantages of using SMTP Proxy, such as a reduction in overall load, an increase in the number of back-end mail servers, and connection management. Furthermore, SMTP Proxy cooperates with the already installed MTA. Another solution is the use of an Anti-Spam SMTP Proxy server [16], which is a transparent SMTP Proxy.

Spam also affects the performance of mail servers due to the unnecessary processing of invalid mails. An efficient solution to this issue is a hybrid mail server architecture that combines the strengths of event-based (efficiency) and process-based (security) architectures [17].

Finally, using anti-spam techniques that employ digitally signed messages (Public Key Infrastructure) may be the most secure, flexible and robust implementation.